\documentclass[review]{elsarticle}
\usepackage{amsfonts}
\usepackage{amsmath}
\usepackage{graphicx}
\usepackage{float}
\usepackage{subfigure}
\usepackage{multirow}
\usepackage{xcolor}
\usepackage{soul}
\usepackage{natbib}
\usepackage{comment}  
\usepackage{amssymb}
\DeclareMathOperator*{\argmax}{argmax}

\begin{document}
\begin{frontmatter}

\title{Contrasting the optimal resource allocation to cybersecurity and cyber insurance using prospect theory versus expected utility theory.}
\author[inst1,inst4]{Chaitanya Joshi\corref{cor1}}
\ead{chaitanya.joshi@auckland.ac.nz}
\author[inst1]{Jinming Yang}
\author[inst2,inst3]{Sergeja Slapni\v{c}ar}
\author[inst2]{Ryan K L Ko}
\cortext[cor1]{Corresponding author}

\affiliation[inst1]{organization={Department of Statistics, University of Auckland},
city={Auckland},
country={New Zealand}}
\affiliation[inst2]{organization={UQ Cyber, University of Queensland},
city={Brisbane}, country={Australia}}
\affiliation[inst3]{organization={UQ Business School, University of Queensland},
city={Brisbane}, country={Australia}}
\affiliation[inst4]{organization={New Zealand Institute for Security and Crime Science}, city={Hamilton},country={New Zealand}}


\begin{abstract}
Protecting against cyber-threats is vital for every organization and can be done by investing in cybersecurity controls and purchasing cyber insurance. However, these are interlinked since insurance premiums could be reduced by investing more in cybersecurity controls. 
The expected utility theory and the prospect theory are two alternative theories explaining decision-making under risk and uncertainty, which can inform strategies for optimizing resource allocation. While the former is considered a rational approach, research has shown that most people make decisions consistent with the latter, including on insurance uptakes. We compare and contrast these two approaches to provide important insights into how the two approaches could lead to different optimal allocations resulting in differing risk exposure as well as financial costs. We introduce the concept of a risk curve and show that identifying the nature of the risk curve is a key step in deriving the optimal resource allocation.\\

\end{abstract}



\end{frontmatter}


\section{Introduction} \label{section:intro}


For the past decade, cyber risk has consistently ranked among the top five business risks, and its significance is only expected to grow \citep[World Economic Forum]{WEF2024}. Effective management of this risk is crucial for every organization. There are two primary approaches to mitigate cyber risk: (i) investing in self-protecting cybersecurity measures (controls) to minimize the likelihood of a cyber incident and (ii) transferring the risk through cyber insurance. Both approaches demand substantial resources, making it essential to optimize the allocation of resources between cybersecurity controls and cyber insurance \citep[]{GordonLoeb2003}. This optimization is particularly challenging due to two key factors.

First, investments in cybersecurity controls and cyber insurance are related \citep[]{Shetty2018}. At a certain level, relying only on cybersecurity controls to reduce risk becomes prohibitively expensive \citep[]{marotta2017survey}, and it is cost-effective to transfer the residual risk to a third party by insuring it. Insurance premiums can be reduced if an organization reduces the insured risk by improving its cybersecurity protection \citep[]{marotta2017survey,UUGAN2021}. Due to interrelated losses and contagion effects of a cyber incident within and across industries, investing in cybersecurity controls improves societal welfare by enhancing the overall level of cybersecurity of organizations and, ultimately, consumers \citep[]{Ogut2011,MUKHOPADHYAY2013,Biener2015,marotta2017survey}. Second, cyber threats are constantly evolving, and new modalities of attacking organizations are being developed, which introduces great uncertainty in cybersecurity decisions \citep{Hubbard2023}. As a result, neither the level of protection cybersecurity controls will provide nor the level of coverage of an insurance can be accurately assessed \citep{Boehm2019}.

Despite the increasing prevalence of cyber risks, cyber insurance remains under-utilized, with demand falling significantly short of expectations given the severity of cyber risks \citep[]{Miller2019,Sullivan2021,MOTT2023}. Insurers face significant difficulties in insuring cyber risks due to the lack of statistical data on incidents, high uncertainty, information asymmetry, the lack of auditing procedures to verify controls of the insured companies, and a small market base to pool risks \citep[]{Yurcik2002,Eling2016,marotta2017survey}. These factors increase insurers’ business risk \citep[]{Eling2016,Shetty2018}, lead to difficult-to-estimate pricing of insurance \citep[]{MOTT2023} and higher premium costs \citep[]{MUKHOPADHYAY2013,Biener2015}. \cite{Eling2019} distinguish between “cyber risks of daily life”, which are insurable, and “extreme cyber risks”, which are not insurable because of diversification properties and changes that render past data less useful. 
 

But even for organizations considering cyber insurance, the diverse consequences of a cyber incident—ranging from financial and operational to managerial, health-related, and reputational—and the interrelated losses make it hard to estimate the potential impact of a cyber incident \citep[]{Ogut2005,Sullivan2021}. Organizations themselves are struggling to quantify their cyber risk financially and tend to assess it qualitatively \citep{Freund2014,Hubbard2023,POLLMEIER2023}. Companies often do not understand well what claims their insurance policy includes \citep[]{Romanosky2016}. Due to high risks that insurers face, policies have unclear coverage, exclusions and limited coverage or low indemnity limits; in short, indemnities tend to be lower than the loss \citep[]{Romanosky2016,marotta2017survey,UUGAN2021}. A risk for an organization is to even prove the loss to the insurance company to get the indemnity. \citep[]{Ogut2011} reported that organizations which incurred a cyber incident could not prove to their insurer the exact amount of a loss to claim the insurance payouts. In the US, the Association of Insurance commissioners found that in 2017 of over 9,000 claims, only $28.4\%$ resulted in a payment by insurance companies \citep{Sullivan2021}. \cite{Cole2021} found that the loss ratio for insurers, that is the ratio between the claims paid out to the insured and premiums collected from them, in the analyzed period from 2014 to 2017 in the USA spanned from only 29 to $51\%$. All these factors might contribute to underinsurance of cyber risks. 


There is an increasing interest in the literature on the underinsurance of severe risks \citep[]{Kunreuther2004,Eling2016,schmidt2016graph}, often referred to as the ‘under-insurance puzzle’ \citep{Browne2015,PITTHAN2021}. This interest sparks inquiries into how individuals make decisions regarding low-probability, high-consequences risks \citep[]{Kunreuther2004,deSmidt2018}, which most people lack experience with. A rational risk-averse decision-maker would, under fair premiums, buy full insurance \citep[]{schmidt2016graph,marotta2017survey}. However, we observe a preference for insuring high-probability, low-consequence risks over low-probability, high consequence ones, even with the same expected loss \citep[]{Sydnor2010,Barseghyan2013,HWANG2021,PITTHAN2021}. Expected utility theory (EUT) fails to explain such behaviors. Consequently, the insurance demand literature has turned to behavioral theories for explanations.\\
To our best knowledge, to date, there are only four studies \citep[]{Ogut2005, marotta2017survey,Rios2021,UUGAN2021} that analyze how an organization optimally allocates the resources to self-protection controls and insurance. All of them assume that the decision-makers make choices that align with EUT. While EUT is seen as a benchmark for rational decision-making, extensive behavioral research documents that prospect theory (PT) \citep[]{kah1979prospect,tversky1992cumulative} can describe the risky choices of the large majority of individuals more accurately \citep[]{PITTHAN2021,Barberis2013}. No published work has analyzed the optimal investment allocation derived from PT. A part of the problem is the application of PT to real-world contexts \citep{Barberis2013}, which is somewhat ambiguous. While the insurance literature has offered numerous behavioral explanations for the insurance demand puzzle, it has yet to propose compelling solutions to this problem.

The objective of this study is to contrast the optimal allocation of funds to cybersecurity controls and cyber insurance under PT to those under EUT. We derive mathematical expressions for the organization's value function under the PT paradigm and the utility function under the EUT paradigm. We consider the following three cases: i) full insurance, ii) under-insurance in which only $80\%$ of the lossis covered, and iii) no insurance, and derive the optimal allocation of resources to cybersecurity controls and cyber insurance (for the first two cases) under both EUT and PT. We uncover some interesting insights into how the two paradigms differ/agree in funds allocation and find that the nature of the risk curve (discussed in Section 4) is crucial to determine which of the two paradigms could be more appropriate.\\
We mathematically show that in the case of full insurance, the allocation of resources to cybersecurity controls under PT will always be less than that under EUT. Our Conjectures illuminate the specific modeling choices in PT that influence the resource allocation. Our numerical examples provide  further insights and show that in case of under-insurance or no insurance, PT will lead to much higher expenditure on cybersecurity controls compared to EUT, but that for most risk curves, this additional expenditure might actually reduce a much higher proportion of risk. Equipped with this knowledge, we illustrate how, for some risk curves, decision-makers who behave in line with PT optimize resource allocation so that they are better protected than if they behaved in line with EUT. With our research, we also address the call by \citep{Eling2021} who, in their systematic literature review, identified a need for research that would develop models to decide on cyber risk treatments while simultaneously considering mitigation controls and insurance. 


The rest of this paper is organized as follows. Section 2 provides the background on EUT and PT as well as literature review of relevant research. In Section 3, we describe the assumptions and derive the mathematical expressions for EUT and PT. The different risk curves and the assumptions are discussed in Section 4, followed by numerical examples in Section 5. Finally, we conclude by discussing the important implications as well as limitations of our results in Section 6.

\section{Background} 

\subsection{EUT} \label{section:EUT}
EUT is a normative account of how individuals (should) make risk choices given their utility function and the initial wealth (von Neumann and Morgenstern, 1947, \cite{VonNeumann1947}). It has been long widely accepted as a standard of rationality \citep{Back2017}. EUT implies that the decision-maker will choose the decision that will maximize the expected value of the utility function. Let $\chi$ be the set of all possible outcomes $x,$ $u(x)$ be the value of the utility function for the outcome $x$ and $p(x)$ be the probability of outcome $x.$ The the expected utility is given by 
\begin{equation}
\label{EUT:eu}
E[u(x)] = \sum_{x\in\chi} u(x)p(x).
\end{equation}

EUT assumes that individuals derive utility from the net wealth, which considers both their initial wealth and the outcomes of their choices. If an individual has initial wealth $W$, and they face a risky decision that could result in additional gains or losses $\pm x$, the resulting net wealth will be $W \pm x$. By explicitly accounting for W, expected utility theory explains that both the magnitude of potential outcomes and their interaction with initial wealth influence decisions.\\
While EUT represents how a rational person would evaluate the outcomes, 
it is possible to model certain risk preferences by using an appropriately formulated utility function. For instance, risk aversion can be modeled in a number of different ways \citep{Back2017}. One of the most commonly used risk aversion formulation \citep{banks2015,Back2017,ejaz21b} is the constant relative risk aversion (CRRA) which is of the form 
\begin{equation}
\label{EUT:crra}
    u(x) =(W-x)^{r},
\end{equation} 
where $x$ represent the cost(loss) associated with a particular choice/action, and $1-r$ is the coefficient of CRRA which measures the proportion of wealth an individual will choose to hold on a risky asset \citep{ejaz21a}.\\
The coefficient $r$ can be used to model a multitude of risk behaviors. $r>1$ represents a risk-seeking (gambling) behavior, $r=1$, represents risk-neutral and $0<r<1$ represents a risk-averse behavior. A risk-averse individual will prefer the certainty of small losses (insurance premiums) over the (small) possibility of a much larger loss. However, often people do not make decisions that are consistent with the EUT \citep{kah1979prospect,tversky1992cumulative,Etner2012,Gilboa2013}. For instance, EUT cannot explain the low uptake of some low probability-high consequences insurance products \citep{deSmidt2018,HWANG2021}.

\subsection{PT} \label{section:PT}

In their two seminal studies about Prospect Theory \citep{kah1979prospect,tversky1992cumulative}, Kahneman and Tversky (henceforth KT) proposed a theory that more accurately describes individuals’ behavior in risky decision-making \footnote{Large studies \citep{bruhin2010} have estimated that up to $80\%$ of people make decisions that are consistent with PT}. PT has four distinct characteristics (KT): (i) People make decisions between risky options not based on the impact on their wealth (as in EUT) but rather in relation to a reference point that leads them to see some outcomes as gains and others as losses (\emph{reference dependence}). (ii) They feel the pain of losses more intensely than the pleasure of an equivalent gain, resulting in loss aversion (\emph{loss aversion}). (iii) The sensitivity to differences in outcomes diminishes with increasing amounts (\emph{diminishing sensitivity}), which renders the value function concave for gains and convex for losses. (iv) Individuals weight probabilities of gains and losses differently and tend to overweight low and underweight high probabilities (\emph{probability weighting}).



From a mathematical perspective, PT is distinct from EUT in how it defines the value function and the probability weights. For the relative (reference dependent) gain/loss $x,$ the \textit{value function} in PT is defined as
\begin{equation}
\label{PT:value}
v(x)=
\left\{
\begin{array}{c}
\ \ \ \ \ x^\alpha ,\ \text{if}\ x\geq0,\\
-\lambda |x|^\alpha ,\ \text{if}\ x<0,
\end{array}
\right.
\end{equation}
where parameter $\alpha$ controls the curvature of the value function and $\lambda$ represents the level of loss aversion. The diminishing sensitivity assumption requires to make the value function concave in the gain domain and convex in the loss domain, this is achieved by restricting the value of $\alpha$ to be between $0$ and $1.$  If the parameter $\lambda=1$, it represents a loss-neutral behavior, whereas, $\lambda>1$ represents loss aversion by giving a higher weight to losses. 
The probability weighting function is defined as
\begin{equation}
\label{PT:weights}
w(p)=\frac{p^\beta}{(p^\beta+(1-p)^\beta)^{1/\beta}},
\end{equation}
where $p$ is the probability and $\beta>0$ is the parameter. When $\beta=1$, decision weights are linear, just like in EUT, where $w(p)=p$. For $\beta<1$, the weight function will be an inverse-S shaped function which will overweight low probabilities and underweight high probabilities. Empirical studies have shown that, $0.5\leq\alpha\leq1$, $1\leq\lambda\leq2.5$ and $0.5\leq\beta\leq1$ \citep{GONZALEZ99,schmidt2008third,bruhin2010}. The median of $\alpha$, $\lambda$ and $\beta$ are 0.88, 2.25 and 0.65 according to \cite{tversky1992cumulative}.\\
These characteristics of PT lead to four patterns of risk choices. While on one hand, individuals tend to be risk averse for gains and risk seeking for losses of high probability, on the other hand, they are risk-seeking for gains and risk averse to losses of low probability \citep{tversky1992cumulative}. The latter is of much relevance to cyber insurance. Individuals are more motivated to avoid further losses than to achieve gains, even if these choices are more risky. This explains why people would prefer a certain small loss to a very small chance of incurring a much larger loss (i.e., why they buy insurance).\\
According to PT, the decision-maker will choose the option that maximizes the overall value $V$ given by 
\begin{equation} \label{PT:valuefn}
    V(x) = \sum_{x\in\chi}v(x)w(p(x)).
\end{equation}

Although PT is currently one of the most prominent descriptive theories of decision-making under risk, a rather limited number of studies apply it to insurance demand \citep{Sydnor2010,Barseghyan2013,schmidt2016graph,HWANG2021}. These studies have considered home, health, long-term care, disability or auto insurances. One of the challenges in applying PT is that it is far from straightforward what a reference point is in different contexts and decisions. A reference point has significant implications for the relationship between loss aversion and the up-take of insurance \cite{HWANG2021}: it determines whether the outcome of a decision is perceived as a gain or a loss and how each of them is weighted.\\
In the original PT, the reference point is the status quo. Prior research has adopted various solutions: \cite{Sydnor2010} and \cite{HWANG2021} interpreted the status quo by considering the initial wealth before the insurance decision is taken and with the loss (that is, adverse event) not occurring. Consequently, the decision to take up insurance is then determined entirely in the loss domain \citep{Sydnor2010}.\\ 
Cyber insurance is not a personal decision for an individual but a decision made by professionals and senior managers for a given organization. This suggests that decision-makers are familiar with the insurance product and  feel that they are capable of estimating cyber risks \citep{Kunreuther2004,deSmidt2018}. For simplicity, we will assume no conflict of interest; that is the problem that the agents do not adopt the decision on their behalf but on behalf of an organization.

\subsection{Prior literature on optimizing allocation of resources into cybersecurity controls and insurance} \label{section:lit_review_EUT}

To the best of our knowledge, only a handful of studies investigate the optimization of resource allocation to cybersecurity controls and insurance. While they differ in the mathematical frameworks, they all use the normative EUT approach for decision-making. 

One of the first such studies was by Ogut et al. (2005) who analyzed the optimal allocation of resources to self-protecting controls and insurance considering the interdependence of cyber risks (when a cyber incident in one organization spills over to others). They focus on analyzing how an optimal decision considering EUT leads to different levels of investing in cybersecurity in an immature market versus a mature market. But it is important to note that the scale and extent of cybersecurity threats as well as the nature of their consequences today is very different from what it was about 20 years ago when this study was conducted. 

More recently, \cite{marotta2017survey} provide a very good overview of the cyber insurance and cybersecurity controls space and lay out a general utility theory framework for optimizing the resource allocation. They 
argue that a risk-averse agent who chooses a certain security level by investing in cybersecurity controls has a higher utility by paying a fixed insurance premium over bearing a random loss. The study also shows that the utility depends on whether the premium is fair, that is, whether it reflects the loss probability and expected loss. However, they do not analytically derive the optimal solution of allocating resources to the two alternative cyber risk mitigation strategies: self-protective controls vs. insurance. Further, finding the optimum insurance premium is considered to be a game between the organization (or their agent) and the insurer, and as such, they use a game theoretic approach to find the optimum premium. But a game theoretic approach does not guarantee that the insurer is able to find the optimal resource allocation for themselves. Also, they do not consider PT at all.

Game theoretic models for cybersecurity resource allocation \citep[]{Cavus2008,rao2016} have been proposed earlier too. But these models assume common knowledge which is especially unrealistic in a cyber insurance setting since information asymmetry is well known \cite{marotta2017survey}. \cite{Rios2021} proposed an adversarial risk analysis (ARA) framework instead that does not make common knowledge assumptions and solves the problem solely from the organization's point of view. They combine ARA with optimization to find the optimal resource allocation for cybersecurity controls and cyber insurance. However, they do not explicitly consider the interdependence between cybersecurity controls and cyber insurance, do not consider PT and the ARA model is for a game between the organization and a specific malicious actor with intent, which limits the overall applicability of their approach.

Not taking the game theoretic view, other studies have considered one of the mathematical optimization approaches instead. \cite{Chase2019} develop a stochastic optimization approach to joint cyber insurance and security-as-a-service provisioning in cloud computing. While their stochastic optimization algorithm is promising, they do not consider the interplay between the insurance costs and the security costs and do not consider any behavioral decision-making aspect including even the risk aversion.

Finally, \cite{UUGAN2021} propose a dynamic programming approach for optimal allocation of resources to different cybersecurity controls and cyber insurance. Their approach also uses the EUT framework set out in \cite{marotta2017survey} but only considers a competitive market with fair premiums and focus on resource allocation into various security controls that are considered independent to each other. However, it too does not consider the loss aversion aspects modeled using PT.

\section{Optimal allocation of resources}

In this section, we describe the set up and derive the optimal allocation of resources to cybersecurity controls and cyber insurance from EUT and PT perspectives. An organization will likely face multiple cyber attacks over a period of time, each of which could result in losses. When an organization considers resource allocation to cybersecurity controls and cyber insurance, they are expecting to protect against all possible attacks and losses. In the set-up below, for the ease of notation and description, we refer to \emph{a successful cyber attack} and \emph{the loss} occuring from a successful cyber attack, however, these are to be interpreted as \emph{at least one successful cyber attack} and the \emph{collective loss} incurred from them, respectively.

Table \ref{table:notation} summarizes the notation detailed in this Section.

\subsection{Set-up}
Let $W$ denote the initial wealth of the organization before investing in cybersecurity controls or cyber insurance and before any losses from a successful cyber attack. Let $L$ denote the loss from a successful cyber attack and $D,$ the indemnity of the insurance cover considered. Let $i_r=D/L$ be the insurance coverage ratio so that $i_r=1$ indicates full insurance, $i_r=0$ indicates no insurance, $i_r>1$ denotes over-insurance and $i_r<1$ denotes under-insurance. As discussed in Section \ref{section:intro} the case of over-insurance for cybersecurity is not realistic and therefore we exclude that case from further consideration. 

Let $C_i$ and $C_{cs}$ denote the cost of cyber insurance and the cost of cybersecurity controls, respectively. We assume that investing in cybersecurity controls reduces the probability of a successful cyber attack and therefore we denote this probability as $\pi(C_{cs})$ for a given $C_{cs}.$ How this probability changes over values of $C_{cs}$ defines, what we refer to as, a \emph{risk curve}. The risk curve plays a key role in deriving the optimal allocation appropriate for a given situation and this is discussed in detail in Section \ref{Section:risk_curves}. We assume that the insurance company will assess the cyber risk using $\pi(C_{cs})$ and derive the cost of the premium accordingly. 

\begin{table}[H]
    \centering
    \begin{tabular}{c|c}
    \hline
       $v(x)$  & value function of the organization, $x$ is the net gain or loss \\
       $w(p)$ & probability weighting function  \\
        $C_{cs}$ & cost of investing in cybersecurity controls\\
       $\pi(C_{cs})$  & probability of a successful cyber attack for a given cybersecurity controls cost \\
        $C_i$ &  cost of cyber insurance (insurance premium)\\
        $C_{tot}$ & total cost $C_{cs}+C_i$\\
        $L$ &  loss due to a successful cyber attack\\
        $W$ &  initial wealth \\
        $i_r$ & insurance coverage ratio, $i_r=1$ for a full coverage\\
        $D$ &  indemnity, $D=i_r L$\\
        $U$ & utility function \\
        \hline
    \end{tabular}
    \caption{Notations}
    \label{table:notation}
\end{table}

In a competitive market where the premiums are fair, we have \citep{marotta2017survey}
\begin{equation}\label{Ci-1}
C_i = \pi(C_{cs}) D = \pi(C_{cs}) i_rL.
\end{equation}
More generally though, an insurance premium will include a profit margin $q,$ where $q$ indicates the rate of profit. For example, $q=0.05$ indicates a $5\%$ profit margin.  To date, there is no fixed standard for pricing cyber insurance, but as an example, insurance companies generally calculate the expected premium (fair) and add a rate $q$ to make a profit. The cost of insurance is then given by
\begin{equation}\label{Ci-2}
C_i = (1+q)\pi(C_{cs}) D = (1+q)\pi(C_{cs}) i_rL.
\end{equation}

Note that the fair premium equation is a special case of Equation \eqref{Ci-2} when $q=0.$ Equations \eqref{Ci-1} and \eqref{Ci-2} make it explicitly clear that $C_i$ is a function of $C_{cs}$ and therefore, the problem of finding optimal values of both $C_i$ and $C_{cs}$ reduces to finding the optimal $C_{cs}$ given $i_r, L, q$ and $\pi(C_{cs}).$ The optimal value of $C_{cs}$ will automatically determine the optimal value $C_i$ using either of the two equations.

\subsection{Prospect Theory solution}

In our analysis of the optimal allocation of resources between cybersecurity controls and insurance, we will adopt the initial wealth (or status quo) as a reference point \citep{Sydnor2010,HWANG2021}. Our decision is based on the belief that individuals code losses and gains ex-ante and that the uncertainty of the state of the world at the time of the decision is fundamental to risky decision-making, especially related to cyber risk. 
The status quo is the standard assumption in PT and is formalized in the third generation of PT \citep{schmidt2008third}. Moreover, wealth before insurance is also justified from the perspective of choosing a ‘social reference point’ based on comparison with others (who generally are not insured) \citep{deSmidt2018}.\\
When using this baseline a decision-maker will always be in the loss domain. In the event of a successful cyber attack, the net loss to an organization is $-C_i - C_{cs}$ if they had full insurance, $-(1-i_r)L -C_i - C_{cs}$ if they had under-insurance  and $-L- C_{cs}$ if they had no insurance. In the event of no successful cyber attack, the net loss will be $-C_i - C_{cs},$ for both the full and under insurance cases and $- C_{cs}$ for the no insurance case. 

The net losses for different scenarios are summarized in Table \ref{table:net_losses_PT}

\begin{table}[H]
    \centering
    \begin{tabular}{|c|c|c|c|}
    \hline
       Successful attack? & Full insurance & Under-insurance & No insurance \\
       \hline
       Yes & $-C_i - C_{cs}$ & $-(1-i_r)L -C_i - C_{cs}$ & $-L- C_{cs}$ \\
       No & $-C_i - C_{cs}$ & $-C_i - C_{cs}$ & $- C_{cs}$\\
    \hline
    \end{tabular}
    \caption{Net losses relative to $W$ under different scenarios for PT}
    \label{table:net_losses_PT}
\end{table}


Using Table \ref{table:net_losses_PT} and Equations \eqref{PT:value}, \eqref{PT:weights}, \eqref{PT:valuefn} and \eqref{Ci-2} the overall value function of the organization if they decide to buy cyber insurance is be given by

\begin{eqnarray}
\label{PT:ir}
\nonumber
V &=& w(\pi(C_{cs}))v(-(1-i_r)L-C_i-C_{cs})+w(1-\pi(C_{cs}))v(-C_{i}-C_{cs}),\\
\nonumber &=& -\lambda \{w(\pi(C_{cs}))\left[L-i_rL(1-(1+q)\pi(C_{cs}))+C_{cs}\right]^\alpha \\
& & + w(1-\pi(C_{cs}))[(1+q)\pi(C_{cs})i_r L + C_{cs}]^\alpha\}.
\end{eqnarray}

In case of full insurance ($i_r=1$) the value function simplifies to
\begin{eqnarray}
\label{PT:ir1}
V & = & -\lambda\left[(1+q)\pi(C_{cs})L + C_{cs}\right]^{\alpha}(w(\pi(C_{cs}))+w(1-\pi(C_{cs}))).
\end{eqnarray}

If the organization did not buy insurance ($i_r=0$) then 
\begin{eqnarray} \label{PT:ir0}
\nonumber  V &=& -\lambda\{ w(\pi(C_{cs}))|-L-C_{cs}|^\alpha+w(1-\pi(C_{cs}))|-C_{cs}|^\alpha \},\\
&=& -\lambda\{ w(\pi(C_{cs}))[L+C_{cs}]^\alpha+w(1-\pi(C_{cs}))[C_{cs}]^\alpha \}
\end{eqnarray}

The optimal solution for resource allocation can be found by first finding $C_{cs}^{*}$ that maximizes the overall value function $V$ and then finding the corresponding $C_{i}^{*}$ using Equation \eqref{Ci-2}.

\subsection{Expected Utility Theory solution}

EUT considers the effect of a decision on net wealth after accounting for any gains/losses. These  are summarized in Table \ref{table:net_losses_EUT} for different scenarios.

\begin{table}[H]
    \centering
    \begin{tabular}{|c|c|c|c|}
    \hline
       Successful attack? & Full insurance & Under-insurance & No insurance \\
       \hline
       Yes & $W-C_i - C_{cs}$ & $W-(1-i_r)L -C_i - C_{cs}$ & $W-L- C_{cs}$ \\
       No & $W-C_i - C_{cs}$ & $W-C_i - C_{cs}$ & $W- C_{cs}$\\
    \hline
    \end{tabular}
    \caption{Net wealth after accounting  for losses under different scenarios for EUT}
    \label{table:net_losses_EUT}
\end{table}

Using Table \ref{table:net_losses_EUT}, the CRRA utility function \eqref{EUT:crra}, and Equations \eqref{EUT:eu} and \eqref{Ci-2} the expected utility function of the organization if it decides to buy cyber insurance is given by
\begin{eqnarray}
\label{EUT-1ir}
\nonumber E[U]&=&\pi(C_{cs})(W-(1-i_r)L-C_i-C_{cs})^{r}+(1-\pi(C_{cs}))(W-C_i-C_{cs})^{r},\\
\nonumber &=& \pi(C_{cs})(W-L-i_rL(1-(1+q)\pi(C_{cs}))-C_{cs})^{r}\\
& & + (1-\pi(C_{cs}))(W-(1+q)\pi(C_{cs})i_rL-C_{cs})^{r}.
\end{eqnarray}

In case of full insurance ($i_r=1$) the function simplifies to
\begin{eqnarray}
\label{EUT-1ir1}
\nonumber E[U]&=&\pi(C_{cs})(W-C_i-C_{cs})^{r}+(1-\pi(C_{cs}))(W-C_i-C_{cs})^{r},\\
&=& (W-(1+q)\pi(C_{cs})L-C_{cs})^{r}.
\end{eqnarray}
If the organization did not buy insurance ($i_r=0$) then
\begin{eqnarray}
\label{EUT-1ir0}
E[U]&=&\pi(C_{cs})(W-L-C_{cs})^{r}+(1-\pi(C_{cs}))(W-C_{cs})^{r}.
\end{eqnarray}

Now, we consider the special case of risk neutral decision-maker, where $r=1$. A risk-neutral attitude indicates that the preferred decision will be solely based on superior expected returns and will not be weighed by the risk involved. 
For this approach, if the organization buys cyber insurance, the expected utility function simplifies to
\begin{eqnarray}
\label{EUT-2ir}
\nonumber
    E[U]&=&\pi(C_{cs})(W-(1-i_r)L-C_i-C_{cs})+(1-\pi(C_{cs}))(W-C_i-C_{cs}),\\
\nonumber    &=& W - \pi(C_{cs})(1-i_r)L-C_i-C_{cs},\\
\nonumber &=& W - \pi(C_{cs})(1-i_r)L-(1+q)\pi(C_{cs})i_rL-C_{cs},\\
&=& W - \pi(C_{cs})L(1+qi_r) - C_{cs}.
\end{eqnarray}
In case of full insurance ($i_r=1$) the expected utility becomes
\begin{eqnarray}
\label{EUT-2ir1}
\nonumber
    E[U]&=& W - \pi(C_{cs})L(1+q) - C_{cs}.
\end{eqnarray}
If the organization did not buy insurance ($i_r=0$) then
\begin{equation}
\label{EUT-2ir0}
    E[U]=W-C_{cs}-\pi(C_{cs})L.
\end{equation}

A unique consequence of the rational behavior modeled here is that, if the organization buys insurance with a fair premium, which means $q=0$, the expected utility function \eqref{EUT-2ir} reduces to \eqref{EUT-2ir0}, when it does not buy the insurance. This makes perfect sense since, by definition, the cost of the fair premium is exactly equal to the expected loss due to an attack in the event of no insurance.\\
The optimal solution for resource allocation can be found by first finding $C_{cs}^{*}$ that maximizes the expected utility and then finding the corresponding $C_{i}^{*}$ using Equation \eqref{Ci-2}.

\subsection{Comparing EUT and PT solutions}

It can be seen from Equations \eqref{EUT:crra} and \eqref{PT:value} that both the utility function and the value function are similar in their functional forms. The utility function $(W-x)^{r}$ is a polynomial of degree $r$. The value function for PT is a polynomial of degree $\alpha$ since it can be expressed as $-\lambda x^{\alpha},$ where $\lambda = -1,$ if $x>0$ and $x=|x|,$ if $x<0$. This implies that both the utility function and the value function are ultimately monotonic functions of $x,$ and as such, the parameters $\lambda, \alpha$ and $r$ should have no effect on the stationary points of the respective functions if we try to optimize these functions. However, the optimal allocation is given by the stationary point of the expected utility function \eqref{EUT:eu} for EUT and the stationary point of the overall value function \eqref{PT:valuefn} for PT. What can we then say about EUT and PT solutions?
  
\subsubsection{General case}

Firstly, from Equations \eqref{PT:ir}, \eqref{PT:ir1} and \eqref{PT:ir0}, it is clear that the loss aversion parameter $\lambda$ will have no effect on the optimal value of $C_{cs}.$ For parameters $r$ we can postulate the following. 

\noindent \textit{\textbf{Conjecture 1:}} The risk aversion parameter $r,$ by itself, will have a minor effect, if any, on the stationary point of the expected utility function.  
\\
Regarding the overall value function V in Equation \eqref{PT:ir} note that as the insurance coverage goes down, $i_r \downarrow,$ the loss in case of a successful attack goes up, $L-i_rL(1-(1+q)\pi(C_{cs}))+C_{cs} \uparrow$ and the cost in case of no such attack goes down, $(1+q)\pi(C_{cs})i_r L + C_{cs} \downarrow.$ The weighting function over-weights small probabilities and under-weights higher ones. Typically, we would expect $\pi(C_{cs})$ to be small and therefore $1-\pi(C_{cs})$ to be larger. Thus, it can be seen that as the insurance coverage decreases, the component corresponding to a successful cyber attack $w(\pi(C_{cs}))\left[L-i_rL(1-(1+q)\pi(C_{cs}))+C_{cs}\right]^\alpha$ will have increasing influence on the overall value function and that this influence will further increase as the diminishing sensitivity parameter $\alpha$ increases. Therefore we postulate the following.

\noindent \textit{\textbf{Conjecture 2:}} The diminishing sensitivity parameter $\alpha,$ could have substantial effect on the stationary point of the overall value function, especially as the insurance coverage ratio $i_r$ decreases. For a given $i_r$, as $\alpha$ increases, so also will the optimal value $C_{cs}^{*}$ of cybersecurity controls expenditure.\\
Finally, note that as the probability weighting parameter increases to 1, $\beta \uparrow 1,$ the non-linearity in the probability weights reduces so that when $\beta=1,$ we have $w(p)=p.$ Consequently, as $\beta \uparrow 1,$ the decision weight on $\pi(C_{cs})$ will decrease reducing the influence of the component corresponding to a successful cyber attack $w(\pi(C_{cs}))\left[L-i_rL(1-(1+q)\pi(C_{cs}))+C_{cs}\right]^\alpha.$ Therefore, we postulate the following.

\noindent \textit{\textbf{Conjecture 3:}} The  probability weighting parameter $\beta$ causes higher allocation of resources to $C_{cs}$ in PT relative to EUT. Increasing the value of $\beta$ (upto a maximum of 1) will reduce the  allocation to $C_{cs}$.

We will use numerical examples to investigate if Conjectures 1, 2 and 3 hold under different risk curves. An exception to Conjecture 2 is the case of full insurance ($i_r=1$) since the overall value equation for that case, see Equation \eqref{PT:ir1}, does not include $L.$ We discuss this case in detail below.
\subsubsection{Special case of Full Insurance, $i_r=1$} \label{sec:full insurance}

For the full insurance case, we can mathematically show that the optimal cybersecurity controls expenditure under PT will always be less than that under EUT.  Further, since the higher expenditure in cybersecurity controls is expected to reduce the risk, the insurance premium under PT will be greater than that under EUT. 



First, we'll prove a mathematical fact that will be used in proving the main results.
\\

\textit{\textbf{Proposition 1:}} When $0<p\leq0.5$, $w(p)+w(1-p)$ is a monotonically decreasing function with respect to $p$.\\
\textit{Proof:}
Because $w(p)+w(1-p)=\frac{p^\beta}{(p^\beta+(1-p)^\beta)^{1/\beta}}+\frac{(1-p)^\beta}{(p^\beta+(1-p)^\beta)^{1/\beta}}.$ Let $x=p^\beta+(1-p)^\beta$, obviously $x>1$ when $0<p\leq 0.5$, then $w(p)+w(1-p)=\frac{x}{x^{1/\beta}}=x^{1-1/\beta}$ is a monotonically decreasing function with respect to $x$, which is $\frac{d[ w(p)+w(1-p)]}{d x}<0$. When $\beta<1$ and $0<p\leq0.5$, $\frac{d x}{d p}>0$. By the chain rule, $\frac{d [w(p)+w(1-p)]}{d x}\frac{d x}{d p}<0$, which means $w(p)+w(1-p)$ is a monotonically decreasing function with respect to $p$. 
\\
Let $C_{EUT}=\argmax_{C_{cs}} E[U]$, $C_{PT}=\argmax_{C_{cs}} V$. Define $\pi(C_{cs})$ is a non-increasing function with respect to $C_{cs}$ and for any $C_{cs}$, $0< \pi(C_{cs})\leq0.5$.\\

\noindent \textit{\textbf{Theorem 1:}} When $i_r=1$, if $E[U]$ only has one stationary point, then $C_{PT}<C_{EUT}$.\\
\textit{\textbf{Proof:}} Because $E[U]$ only has one stationary point and $C_{EUT}=\argmax_{C_{cs}} E[U]$, we have
\begin{equation}
\frac{\partial E[U(C_{EUT})]}{\partial C_{cs}}=-1-\pi'(C_{EUT})(1+q)L=0.
\label{theo1 1}
\end{equation}
Consider the value of $\frac{\partial V(C_{EUT})}{\partial C_{cs}}$
\begin{eqnarray}
\nonumber
&=&-\lambda\{\alpha[\pi(C_{EUT})(1+q)L+C_{EUT}]^{\alpha-1}[\pi'(C_{EUT})(1+q)L+1][w(\pi(C_{EUT}))+w(1-\pi(C_{EUT}))]\\
&&+[\pi(C_{EUT})(1+q)L+C_{EUT}]^{\alpha}\frac{\partial [w(\pi)+w(1-\pi)]}{\partial \pi}\frac{\partial \pi}{\partial C_{cs}}|_{C_{cs}=C_{EUT}}\}.
\label{theo1 2}
\end{eqnarray}
By \eqref{theo1 1}, the first part of \eqref{theo1 2} is 0, so \eqref{theo1 2} can be simplified to
\begin{equation}
\nonumber
\frac{\partial V(C_{EUT})}{\partial C_{cs}}=-\lambda\{ [\pi(C_{EUT})(1+q)L+C_{EUT}]^{\alpha}\frac{\partial [w(\pi)+w(1-\pi)]}{\partial \pi}\frac{\partial \pi}{\partial C_{cs}}|_{C_{cs}=C_{EUT}}\}.
\end{equation}
By \textit{Proposition 1}, and for any $C_{cs}$, $\frac{\partial \pi}{\partial C_{cs}}\leq0$, we know that $\frac{\partial [w(\pi)+w(1-\pi)]}{\partial \pi}\frac{\partial \pi}{\partial C_{cs}}|_{C_{cs}=C_{EUT}}\ \geq 0.$ But \eqref{theo1 1} shows that $\pi'(C_{EUT})\ne0$, so $\frac{\partial [w(\pi)+w(1-\pi)]}{\partial \pi}\frac{\partial \pi}{\partial C_{cs}}|_{C_{cs}=C_{EUT}}\}> 0.$ Since $[\pi(C_{EUT})(1+q)L+C_{EUT}]^{\alpha}>0$, we have $\frac{\partial V(C_{EUT})}{\partial C_{cs}}<0$, which means $V$ is already decreasing at $C_{EUT}$. So
\begin{equation}
\nonumber
C_{PT}=\argmax_{C_{cs}} V<C_{EUT}.
\end{equation}

\noindent \textit{\textbf{Theorem 2:}} When $i_r=1$, if $E[U]$ only has one stationary point, then $C_{i EUT}< C_{i PT}$.\\
\noindent\textit{\textbf{Proof:}} By \textit{Theorem 1}, we have $C_{PT}<C_{EUT}$. Because 
\begin{equation}
\nonumber
    C_{i EUT}=\pi(C_{EUT})(1+q)L, C_{i PT}=\pi(C_{PT})(1+q)L,
\end{equation}
and $\pi'(C_{EUT})\ne0$, $\pi(C_{cs})$ is a non-increasing function with respect to $C_{cs}$, then $C_{i EUT}< C_{i PT}$.\\

There will be some cases in which the total cost of EUT is equal to PT, which is $C_{EUT}+C_{iEUT}=C_{PT}+C_{iPT}.$ Expanding it we have
\begin{equation}
   C_{EUT}+\pi(C_{EUT})(1+q)L=C_{PT}+\pi(C_{PT})(1+q)L.
\end{equation}

By \textit{Theorem 1} and \textit{Theorem 2}, we know that $C_{ EUT}> C_{ PT}$ and $C_{i EUT}< C_{i PT}$, so if $L$, $q$ and $\pi$ satisfy the following condition
\begin{equation}
    C_{EUT}-C_{PT}=(1+q)L[\pi(C_{PT})-\pi(C_{EUT})],
\end{equation}
then the total cost of EUT is equal to PT.

\section{Risk Curves} \label{Section:risk_curves}

By a risk curve we mean a function that models how the probability of a successful cyber attack changes with the investment in cybersecurity controls.  In this section we use numerical examples to show how the optimal $C_{cs},$ and therefore optimal $C_{i},$ changes for differently shaped risk curves (which represent different scenarios).\\
Let $\pi$ denote the probability of (at least one) successful cyber attack. We make some basic assumptions that we believe are very realistic. These are also consistent with the framework recommended by \cite{marotta2017survey}.
\begin{enumerate}
    \item Investment in cybersecurity controls affects this probability. That is, $\pi = \pi(C_{cs})$ is a function of $C_{cs}.$
    \item $\pi(C_{cs})$ is monotonically non-decreasing, that is, $\pi$ will either decrease or stay constant with increase in $C_{cs}.$
    \item Many software products and IT systems (including network as well as cloud, etc) come with security measures built into them. As a result, the baseline risk $\pi(C_{cs} =0)$ is substantially smaller than 1.
    \item $\pi(C_{cs})>0,$ that is, the probability of (at least one) successful cyber attack can never be completely eliminated.
\end{enumerate}


Here we consider five different risk curves that we believe represent different scenarios that can occur in practice. These are to be considered merely as templates indicating different relationships between $C_{cs}$ and $\pi(C_{cs}).$
\begin{itemize}
    \item {\textit{Slow vs. rapidly decreasing:} Curves $\pi_1$ and $\pi_3$ represent scenarios where the probability reduces slowly with increasing the investment in controls and this rate  slows down with further increasing investment in controls.  These represent situations where reducing cyber risk is challenging and expensive. On the other hand curves $\pi_2$ represents situation where further investment causes a rather rapid decrease in the risk, although, this decline still slows down with increasing investment.}
      \item {\textit{Baseline risk:} Curves $\pi_1$ and $\pi_2$ have a baseline probability of about $0.2.$ This is the probability of (at least one) successful attack if there is no investment in cybersecurity controls (that is, $C_{cs} =0$). On the other hand curve $\pi_3$  represents a situation where the baseline probability is higher at $0.3.$}
    \item \textit{Stepwise reduction:} Curves $\pi_4$ and $\pi_5$ represent situations where the probability does initially decline with the investment in controls, but then plateaus off. Any further reduction in probability requires substantial further investment in controls (in a different type of technology, etc). Thus, the reduction in cyber risk is both challenging and expensive in this scenario. Curve $\pi_5$ represents a situation where there are three such steps.
\end{itemize}

The curves have been modeled using exponential functions of the form $e^{-\lambda C_{cs}},$ where $\lambda>0$ is a constant rate at which $\pi(C_{cs})$ decreases with increasing $C_{cs}$. As a result, these curves are asymptotic and $\pi(C_{cs})>0$ for any $C_{cs}<\infty.$ Curves $\pi_1$ and $\pi_3$ use the same value of $\lambda,$ whereas $\pi_2$ has been modeled using a higher value of $\lambda$ to indicate a rapidly decreasing curve. Stepwise curves use multiple $\lambda$ values. Since we use these curves as mere templates, the actual $\lambda$ values used are not important. In fact, every organization will need to derive a risk curve that is the best fit for them and could be unique to them. 

\begin{figure}[H]
    \centering
    \includegraphics[width = .9\textwidth]{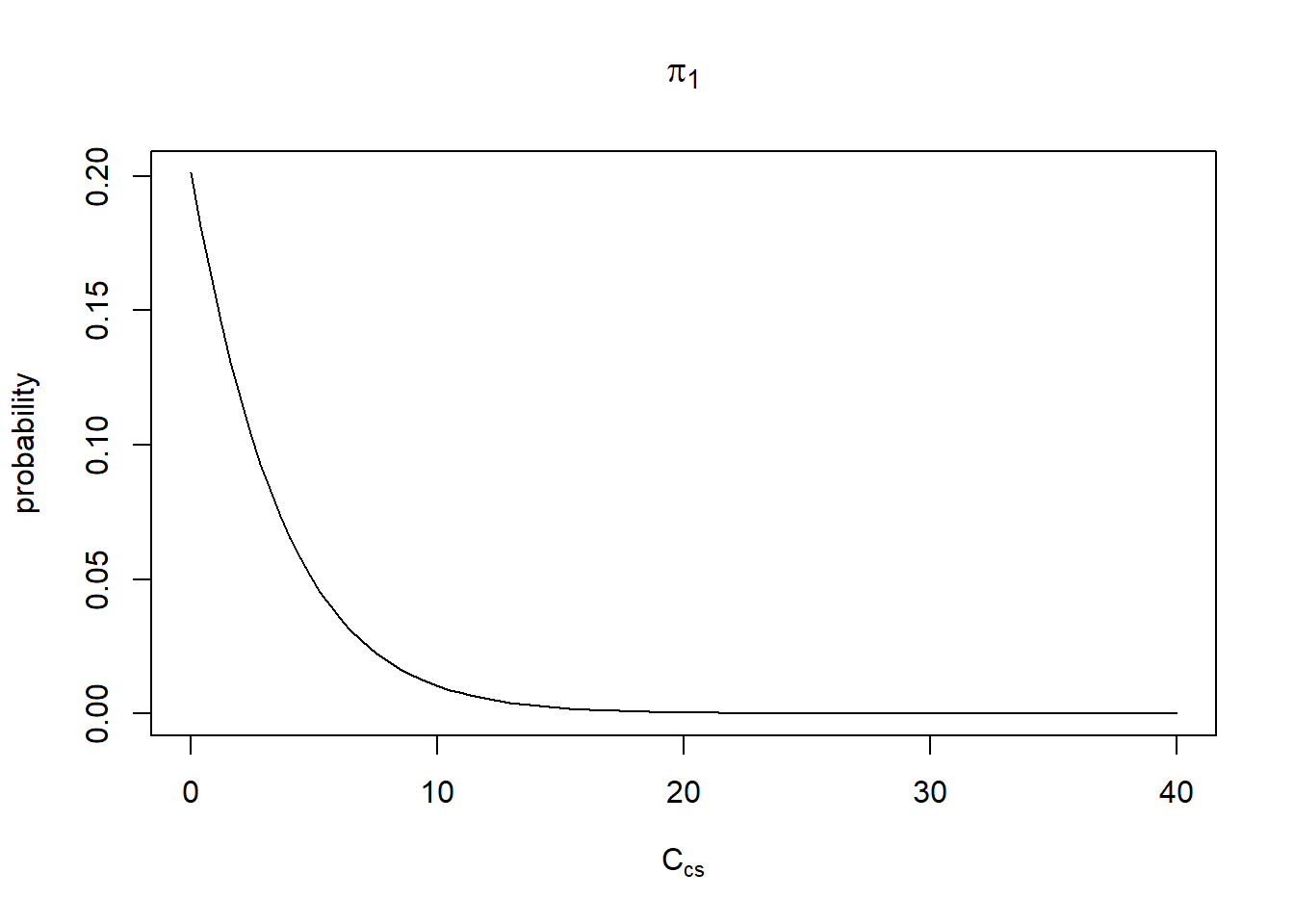}
    \caption{$\pi_1$}
    \label{fig:pi1}
\end{figure}
\begin{figure}[H]
    \centering
    \includegraphics[width = .9\textwidth]{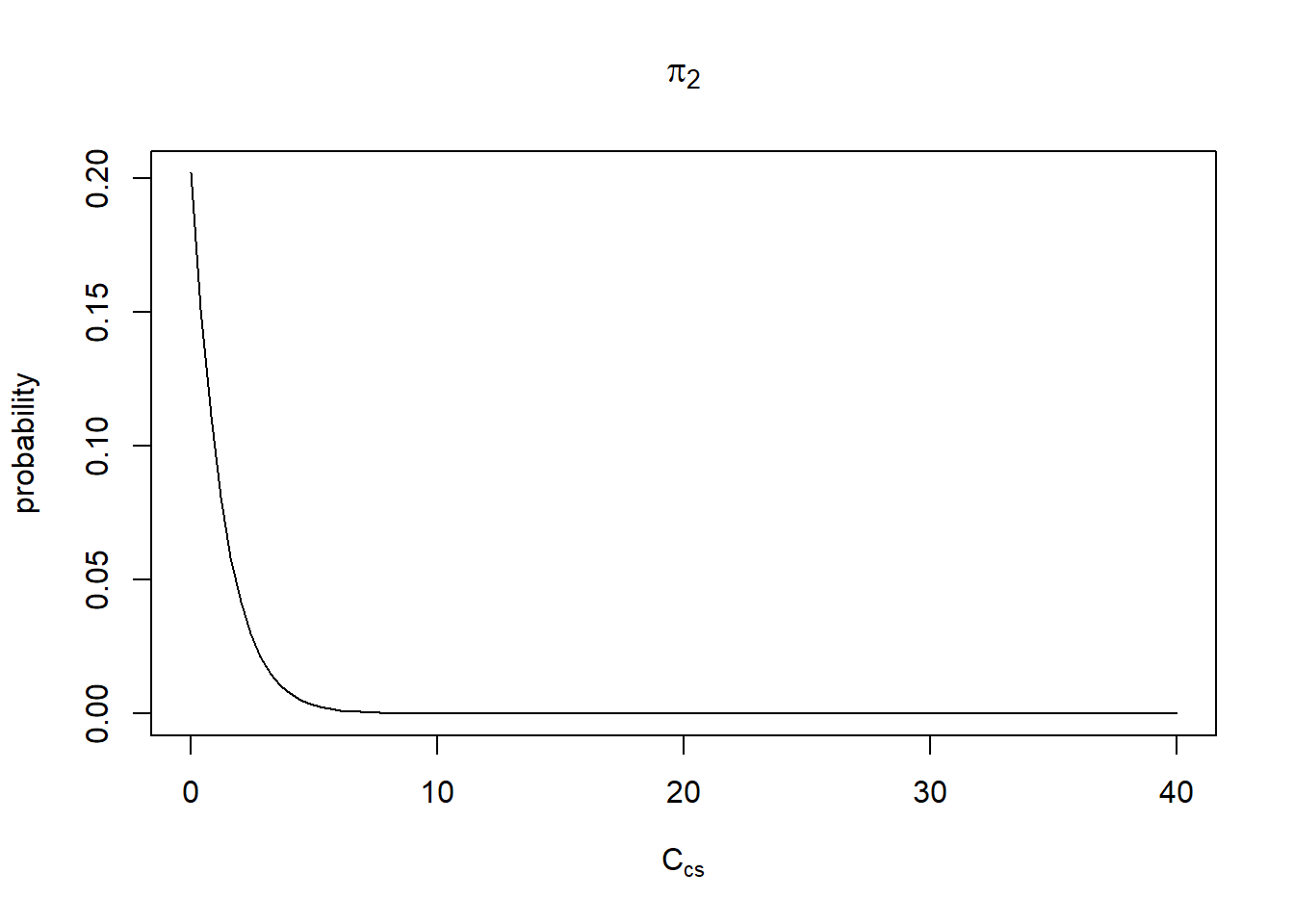}
    \caption{$\pi_2$}
    \label{fig:pi2}
\end{figure}
\begin{figure}[H]
    \centering
    \includegraphics[width = .9\textwidth]{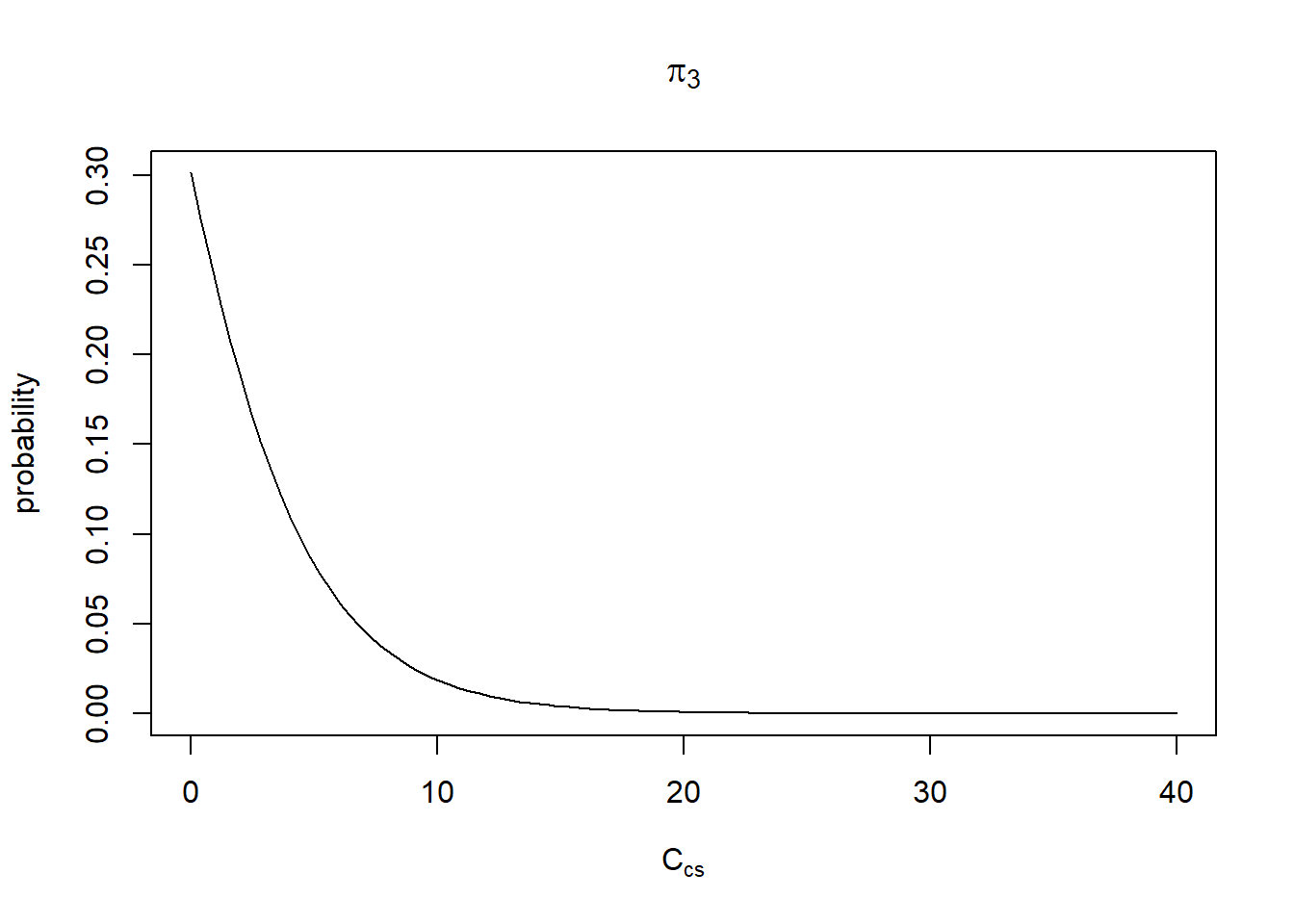}
    \caption{$\pi_3$}
    \label{fig:pi3}
\end{figure}
\begin{figure}[H]
    \centering
    \includegraphics[width = .9\textwidth]{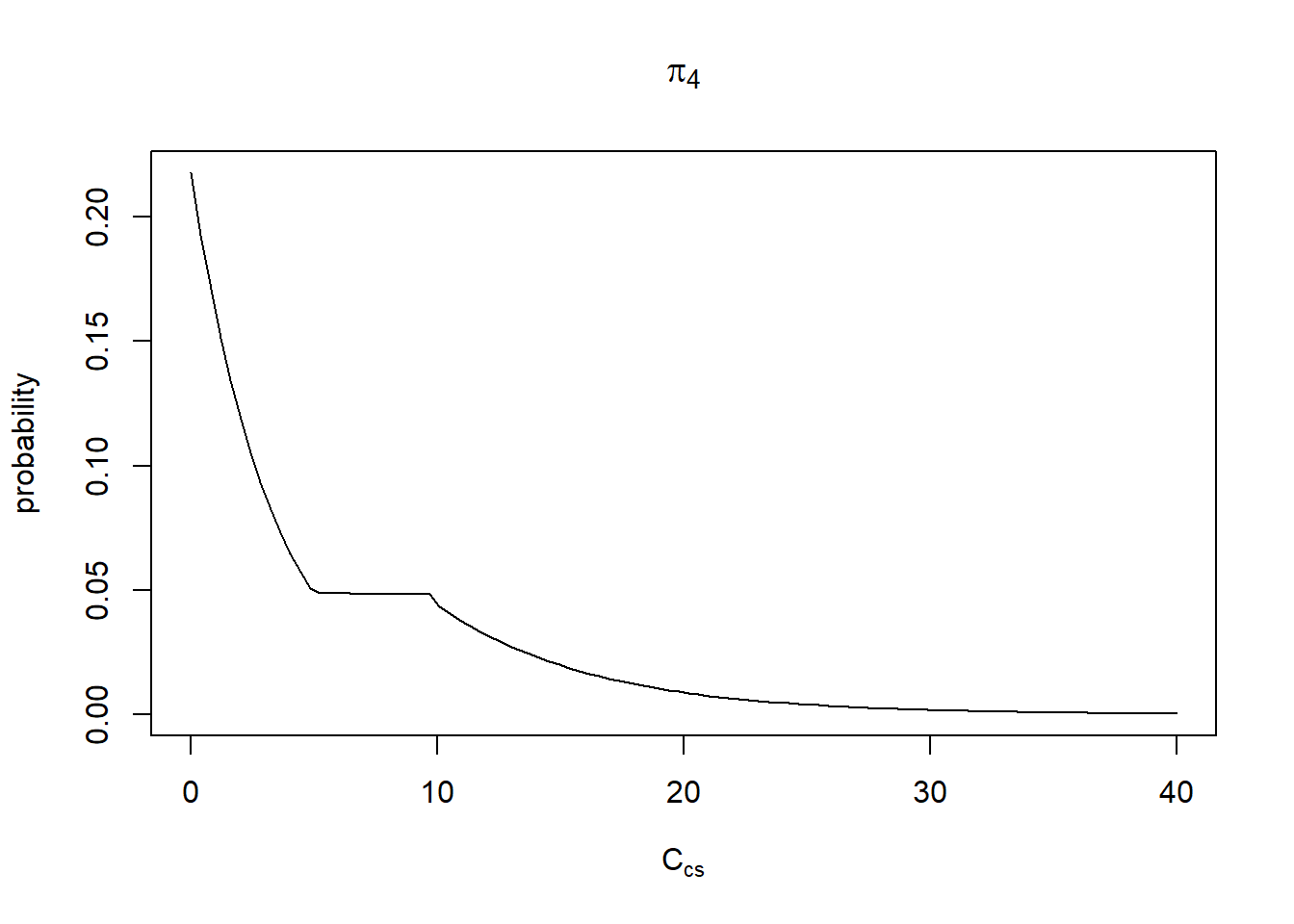}
    \caption{$\pi_4$}
    \label{fig:pi4}
\end{figure}
\begin{figure}[H]
    \centering
    \includegraphics[width = .9\textwidth]{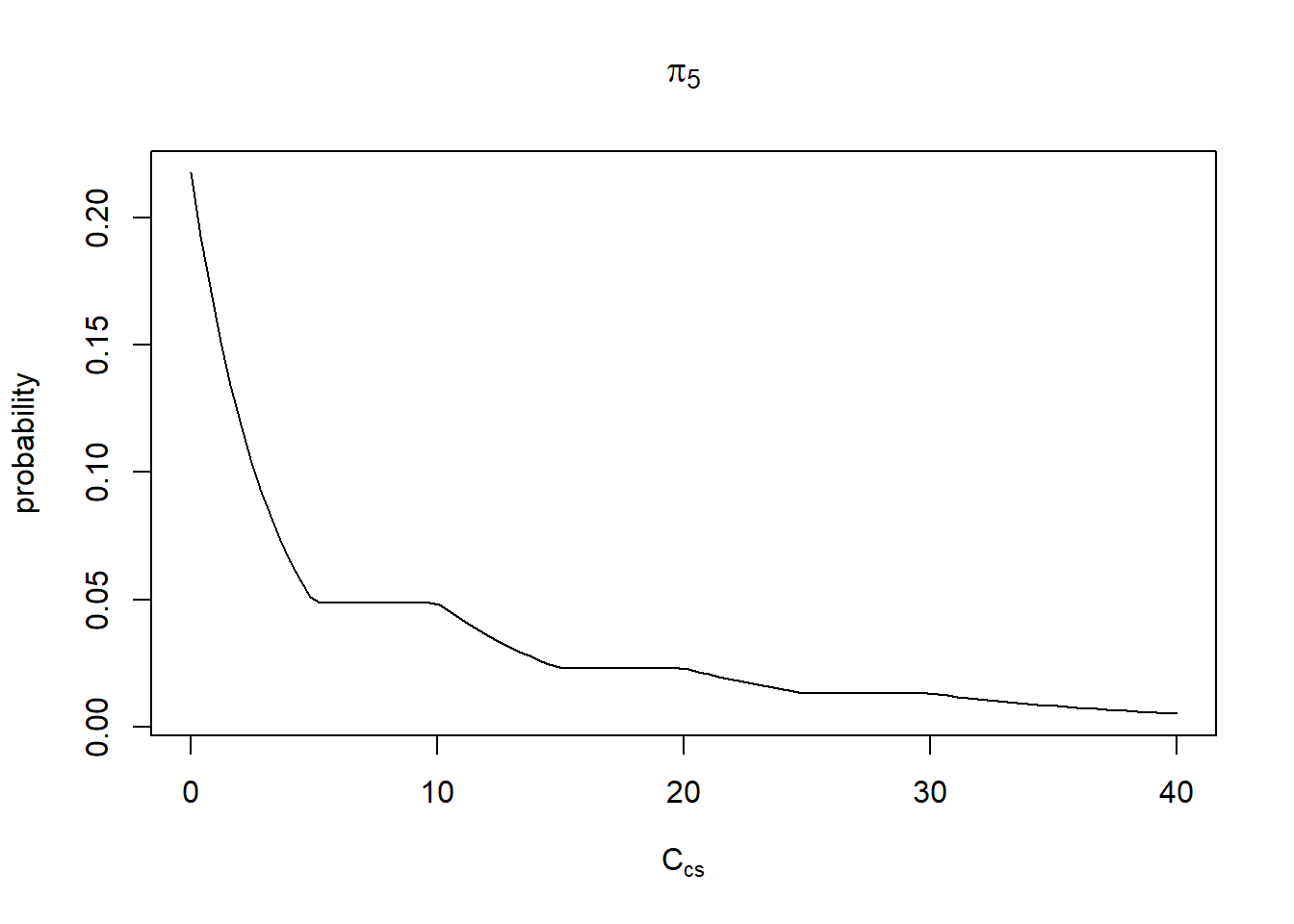}
    \caption{$\pi_5$}
    \label{fig:pi5}
\end{figure}

\section{Numerical comparisons}
We now run numerical comparisons to illustrate how the optimal resource allocation considering the PT and the EUT perspectives might differ for the five different risk curves discussed in Section \ref{Section:risk_curves} and to investigate the three Conjectures. This will provide a better understanding of the modeling inputs that influence the optimal allocation.  We find the optimal solutions for three distinct cases:  full insurance ($i_r=1$), under insurance ($i_r=0.8$) and no insurance ($i_r=0$).\\
We assume that $W=10,000, L =1,000,$ and the profit margin $q=0.3$ ($30\%$). Thus, we find optimal solutions for the general case of non-fair premiums and assume that the losses due to successful cyber attacks are about $10\%$ of the initial wealth.

\subsection{Investigating Conjectures 1, 2 and 3}
For PT, we consider the median values from \cite{tversky1992cumulative}, $\alpha=0.88, \beta=0.65$ and $\lambda=2.25$ as the base case.\\
To investigate Conjecture 1, we find the optimal solutions for two cases: risk neutral ($r=1$) and risk averse. Keeping in mind the similarity in expressions for EUT and PT, we choose the risk aversion parameter $r$ to be equal to the base case value of $\alpha$ in PT. That is, we consider the cases of $r=0.88$ and $r=1.$\\
To investigate Conjecture 2, we consider different values of $\alpha.$ Note that $\alpha =1$ indicates no diminishing sensitivity to the increasing value of losses. \citep{bruhin2010} found that while $\alpha$ could be as low as $0.25$ (indicating a very strong level of diminishing sensitivity) in some experiments, in most cases, it was found to be between $0.5$ and $1.$ We consider the median value of $\alpha$ to be a \emph{moderate} level and consider two PT variations:  \emph{high level of diminishing sensitivity} with $\alpha=0.65$ and \emph{low level of diminishing sensitivity} with $\alpha=0.95.$\\


\begin{table}[H]
    \centering
    \begin{tabular}{|c|c|c|c|c|c|c|c|c|c|}
    \hline
    \multirow{2}{*}{$\pi_1$} &\multicolumn{3}{|c|}{$i_r=0$} &\multicolumn{3}{|c|}{$i_r=0.8$}& \multicolumn{3}{|c|}{$i_r=1$} \\ \cline{2-10}
     & $C_{cs}$ &$C_i$& $C_{tot}$ &$C_{cs}$ &$C_i$& $C_{tot}$&$C_{cs}$ &$C_i$& $C_{tot}$ \\ \hline
    PT, $\alpha=0.65$ &16.16 &0 &16.16 &15.46 &2.2 &17.66 & 14.64 &3.51 &18.16 \\ \hline
    PT, $\alpha=0.88$ &19.98 &0 &19.98 &16.14 &1.79 &17.93 &14.69 &3.46 &18.15 \\ \hline
    PT, $\alpha=0.95$ &21.08 &0 &21.08 &16.39 &1.66 &18.05 &14.7 &3.45 &18.15 \\ \hline
    EUT, $r=0.88$ &13.96 &0 &13.96 &14.66 &2.80 &17.46 &14.82 &3.33 &18.15\\ \hline
    EUT, $r=1$ &13.94 &0 &13.94 &14.66 &2.8 &17.46 &14.82 &3.33 &18.15 \\ \hline
    \end{tabular}
    \caption{Optimal $C_{cs}, C_i$  and $C_{tot}$ for $\pi_1$}
    \label{tab:pi1_1}
\end{table}

\begin{table}[H]
    \centering
    \begin{tabular}{|c|c|c|c|c|c|c|c|c|c|}
    \hline
    \multirow{2}{*}{$\pi_2$} &\multicolumn{3}{|c|}{$i_r=0$} &\multicolumn{3}{|c|}{$i_r=0.8$}& \multicolumn{3}{|c|}{$i_r=1$} \\ \cline{2-10}
     & $C_{cs}$ &$C_i$& $C_{tot}$ &$C_{cs}$ &$C_i$& $C_{tot}$&$C_{cs}$ &$C_i$& $C_{tot}$ \\ \hline
    PT, $\alpha=0.65$ &6.88 &0 &6.88 &6.51 &0.67 &7.18 & 6.16 &1.15 &7.31 \\ \hline
    PT, $\alpha=0.88$ &8.44&0&8.44 &6.88 &0.48 &7.36 &6.17 &1.14 &7.31\\ \hline
    PT, $\alpha=0.95$ &8.58 &0 &8.58 &7.03 &0.42 &7.45 &6.17 &1.14 &7.31 \\ \hline
    EUT, $r=0.88$ &5.91&0&5.91 &6.14 &0.93 &7.08 &6.20 &1.11 &7.31 \\ \hline
    EUT, $r=1$ &5.91&0&5.91 &6.14 &0.93 &7.08 &6.20 &1.11 &7.31 \\ \hline
    \end{tabular}
    \caption{Optimal $C_{cs}, C_i$  and $C_{tot}$ for $\pi_2$}
    \label{tab:pi2_1}
\end{table}

\begin{table}[H]
    \centering
    \begin{tabular}{|c|c|c|c|c|c|c|c|c|c|}
    \hline
    \multirow{2}{*}{$\pi_3$} &\multicolumn{3}{|c|}{$i_r=0$} &\multicolumn{3}{|c|}{$i_r=0.8$}& \multicolumn{3}{|c|}{$i_r=1$} \\ \cline{2-10}
     & $C_{cs}$ &$C_i$& $C_{tot}$ &$C_{cs}$ &$C_i$& $C_{tot}$&$C_{cs}$ &$C_i$& $C_{tot}$ \\ \hline
    PT, $\alpha=0.65$ &18.08 &0 &18.08 &17.23 &2.19 &19.42 & 16.38 &3.53 & 19.91 \\ \hline
    PT, $\alpha=0.88$ &21.78 &0 &21.78 &17.90 &1.79 &19.69 &16.43 &3.48 &19.91\\ \hline
    PT, $\alpha=0.95$ &22.85 &0 &22.85 &18.14 &1.67 &19.81 &16.44 &3.47 &19.91 \\ \hline
    EUT, $r=0.88$ &15.72 &0 &15.72 &16.42 &2.80 &19.21 &16.57 &3.33 &19.91\\ \hline
    EUT, $r=1$ &15.70& 0 &15.70 &16.42 &2.80 &19.21 &16.57 &3.33 &19.91 \\ \hline
    \end{tabular}
    \caption{Optimal $C_{cs}, C_i$  and $C_{tot}$ for $\pi_3$}
    \label{tab:pi3_1}
\end{table}

\begin{table}[H]
    \centering
    \begin{tabular}{|c|c|c|c|c|c|c|c|c|c|}
    \hline
    \multirow{2}{*}{$\pi_4$} &\multicolumn{3}{|c|}{$i_r=0$} &\multicolumn{3}{|c|}{$i_r=0.8$}& \multicolumn{3}{|c|}{$i_r=1$} \\ \cline{2-10}
     & $C_{cs}$ &$C_i$& $C_{tot}$ &$C_{cs}$ &$C_i$& $C_{tot}$&$C_{cs}$ &$C_i$& $C_{tot}$ \\ \hline
    PT, $\alpha=0.65$ &25.32 &0 &25.32 &24.67 &4.39 &29.06 & 23.44 &6.68 &30.12 \\ \hline
    PT, $\alpha=0.88$ &31.71 &0 &31.71 &25.60 &3.78 &29.38 &23.55 &6.57 &30.11\\ \hline
    PT, $\alpha=0.95$ &33.5 &0 &33.5 &25.93 &3.59 &29.52 &23.57 &6.54 &30.11\\ \hline
    EUT, $r=0.88$ &22.25 &0 &22.25 &23.56 &5.24 &28.80 &23.85 &6.25 &30.10\\ \hline
    EUT, $r=1$ &22.21 &0 &22.21 &23.56 &5.24 &28.80 &23.85 &6.25 &30.10 \\ \hline
    \end{tabular}
    \caption{Optimal $C_{cs}, C_i$  and $C_{tot}$ for $\pi_4$}
    \label{tab:pi4_1}
\end{table}

\begin{table}[H]
    \centering
    \begin{tabular}{|c|c|c|c|c|c|c|c|c|c|}
    \hline
    \multirow{2}{*}{$\pi_5$} &\multicolumn{3}{|c|}{$i_r=0$} &\multicolumn{3}{|c|}{$i_r=0.8$}& \multicolumn{3}{|c|}{$i_r=1$} \\ \cline{2-10}
     & $C_{cs}$ &$C_i$& $C_{tot}$ &$C_{cs}$ &$C_i$& $C_{tot}$&$C_{cs}$ &$C_i$& $C_{tot}$ \\ \hline
    PT, $\alpha=0.65$ &15&0&15 &25 &13.65 &38.65 &25 &17.06 &42.06\\ \hline
    PT, $\alpha=0.88$ &25&0&25 &25 &13.65 &38.65 &25 &17.06 &42.06\\ \hline
    PT, $\alpha=0.95$ &25&0&25 &25 &13.65 &38.65 &25 &17.06 &42.06\\ \hline
    EUT, $r=0.88$ &25&0&25 &25 &13.65 &38.65 &25 &17.06 &42.06\\ \hline
    EUT, $r=1$ &25&0&25 &25 &13.65 &38.65 &25 &17.06 &42.06 \\ \hline
    \end{tabular}
    \caption{Optimal $C_{cs}, C_i$  and $C_{tot}$ for $\pi_5$}
    \label{tab:pi5_1}
\end{table}

The results in Tables \ref{tab:pi1_1} to \ref{tab:pi5_1} provide clear evidence that Conjecture 1 could be true. That is, regardless of the risk curve and the level of insurance, the optimal resource allocation of investments in cybersecurity controls and cyber insurance considering the EUT perspective only changes a little, if at all,  with change in  the risk aversion coefficient $r.$ \\
Except for the risk curve $\pi_5,$ the allocation to $C_{cs}$ under the PT perspective seems to increase as $\alpha$ increases and especially so, as $i_r$ decreases. These results provide a clear indication that Conjecture 2 could be true. That is, the diminishing sensitivity parameter $\alpha$ has a substantial effect on the stationary point of the overall value function, especially as the insurance coverage ratio $i_r$ decreases.\\
To investigate Conjecture 3, we now repeat the PT allocations for the case of linear weights ($\beta=1.$). 

\begin{table}[H]
    \centering
    \begin{tabular}{|c|c|c|c|c|c|c|c|c|c|}
    \hline
    \multirow{2}{*}{$\pi_1$} &\multicolumn{3}{|c|}{$i_r=0$} &\multicolumn{3}{|c|}{$i_r=0.8$}& \multicolumn{3}{|c|}{$i_r=1$} \\ \cline{2-10}
     & $C_{cs}$ &$C_i$& $C_{tot}$ &$C_{cs}$ &$C_i$& $C_{tot}$&$C_{cs}$ &$C_i$& $C_{tot}$ \\ \hline
    PT, $\alpha=0.65$ &9.87 &0 &9.87 &14.41 &3.01 &17.42 & 14.82 &3.33 &18.15 \\ \hline
    PT, $\alpha=0.88$ &12.59 &0 &12.59 &14.56 &2.88 &17.44 &14.82 &3.33 &18.15 \\ \hline
    PT, $\alpha=0.95$ &13.38 &0 &13.38 &14.61 &2.83 &17.45 &14.82 &3.33 &18.15 \\ \hline
    \end{tabular}
    \caption{Optimal $C_{cs}, C_i$  and $C_{tot}$ for the three PT variations for $\pi_1$ when $\beta=1.$}
    \label{tab:pi1_2}
\end{table}

\begin{table}[H]
    \centering
    \begin{tabular}{|c|c|c|c|c|c|c|c|c|c|}
    \hline
    \multirow{2}{*}{$\pi_2$} &\multicolumn{3}{|c|}{$i_r=0$} &\multicolumn{3}{|c|}{$i_r=0.8$}& \multicolumn{3}{|c|}{$i_r=1$} \\ \cline{2-10}
     & $C_{cs}$ &$C_i$& $C_{tot}$ &$C_{cs}$ &$C_i$& $C_{tot}$&$C_{cs}$ &$C_i$& $C_{tot}$ \\ \hline
    PT, $\alpha=0.65$ &4.23 &0 &4.23 &6.04 &1.03 &7.06 & 6.2 &1.11 &7.31 \\ \hline
    PT, $\alpha=0.88$ &5.35 &0 &5.35 &6.1 &0.97 &7.07 &6.2 &1.11 &7.31\\ \hline
    PT, $\alpha=0.95$ &5.67 &0 &5.67 &6.12 &0.95 &7.07 &6.2 &1.11 &7.31 \\ \hline
    \end{tabular}
    \caption{Optimal $C_{cs}, C_i$  and $C_{tot}$ for the three PT variations for $\pi_2$ when $\beta=1.$}
    \label{tab:pi2_2}
\end{table}

\begin{table}[H]
    \centering
    \begin{tabular}{|c|c|c|c|c|c|c|c|c|c|}
    \hline
    \multirow{2}{*}{$\pi_3$} &\multicolumn{3}{|c|}{$i_r=0$} &\multicolumn{3}{|c|}{$i_r=0.8$}& \multicolumn{3}{|c|}{$i_r=1$} \\ \cline{2-10}
     & $C_{cs}$ &$C_i$& $C_{tot}$ &$C_{cs}$ &$C_i$& $C_{tot}$&$C_{cs}$ &$C_i$& $C_{tot}$ \\ \hline
    PT, $\alpha=0.65$ &11.82 &0 &11.82 &16.18 &3.0 &19.18 & 16.57 &3.33 & 19.91 \\ \hline
    PT, $\alpha=0.88$ &14.4 &0 &14.4 &16.32 &2.88 &19.2 &16.57 &3.33 &19.91\\ \hline
    PT, $\alpha=0.95$ &15.16 &0 &15.16 &16.37 &2.83 &19.2 &16.57 &3.33 &19.91 \\ \hline
    \end{tabular}
       \caption{Optimal $C_{cs}, C_i$  and $C_{tot}$ for the three PT variations for $\pi_3$ when $\beta=1.$}
    \label{tab:pi3_2}
\end{table}

\begin{table}[H]
    \centering
    \begin{tabular}{|c|c|c|c|c|c|c|c|c|c|}
    \hline
    \multirow{2}{*}{$\pi_4$} &\multicolumn{3}{|c|}{$i_r=0$} &\multicolumn{3}{|c|}{$i_r=0.8$}& \multicolumn{3}{|c|}{$i_r=1$} \\ \cline{2-10}
     & $C_{cs}$ &$C_i$& $C_{tot}$ &$C_{cs}$ &$C_i$& $C_{tot}$&$C_{cs}$ &$C_i$& $C_{tot}$ \\ \hline
    PT, $\alpha=0.65$ &15.52 &0 &15.52 &23.18 &5.57 &28.75 & 23.85 &6.25 &30.1 \\ \hline
    PT, $\alpha=0.88$ &20.01 &0 &20.01 &23.41 &5.37 &28.78 &23.85 &6.25 &30.1\\ \hline
    PT, $\alpha=0.95$ &21.3 &0 &21.3 &23.5 &5.3 &28.8 &23.85 &6.25 &30.1\\ \hline
    \end{tabular}
   \caption{Optimal $C_{cs}, C_i$  and $C_{tot}$ for the three PT variations for $\pi_4$ when $\beta=1.$}
    \label{tab:pi4_2}
\end{table}

\begin{table}[H]
    \centering
    \begin{tabular}{|c|c|c|c|c|c|c|c|c|c|}
    \hline
    \multirow{2}{*}{$\pi_5$} &\multicolumn{3}{|c|}{$i_r=0$} &\multicolumn{3}{|c|}{$i_r=0.8$}& \multicolumn{3}{|c|}{$i_r=1$} \\ \cline{2-10}
     & $C_{cs}$ &$C_i$& $C_{tot}$ &$C_{cs}$ &$C_i$& $C_{tot}$&$C_{cs}$ &$C_i$& $C_{tot}$ \\ \hline
    PT, $\alpha=0.65$ &15 & 0 &15 &25 &13.65 &38.65 &25 &17.06 &42.06\\ \hline
    PT, $\alpha=0.88$ &15 & 0 &25 &25 &13.65 &38.65 &25 &17.06 &42.06\\ \hline
    PT, $\alpha=0.95$ &15 & 0 &25 &25 &13.65 &38.65 &25 &17.06 &42.06\\ \hline
    \end{tabular}
      \caption{Optimal $C_{cs}, C_i$  and $C_{tot}$ for the three PT variations for $\pi_5$ when $\beta=1.$}
    \label{tab:pi5_2}
\end{table}

Tables \ref{tab:pi1_2} to \ref{tab:pi5_2} show that when the probability weighting is linear, $\beta=1$ (same as in EUT), the optimal resource allocation to $C_{cs}$ under the PT perspective is significantly lower than when $\beta = 0.65$. This is true for all risk curves, except for risk curve $\pi_5,$ where the allocations for the cases of $i_r=0.8$ and $i_r=1$ remain unchanged. Thus, the results indicate that the probability weighting parameter $\beta$ being less than $1$ (non-linear weights) causes higher allocation of resources to $C_{cs}$ under the PT perspective relative to the allocation under the EUT perspective. That is, Conjecture 3 also seems to hold true. 

\subsection{Risk reduction vs. cost optimization}

So far we have seen that except for the case of full insurance, a decision-maker tends to allocate more resources towards $C_{cs}$ if they act as per PT than if they act as per EUT. While this additional expenditure results in lower insurance costs $C_{i},$ the total expenditure $C_{tot}$ under the PT perspective is still higher than under the EUT perspective. Thus, the EUT perspective seems to lead to a resource allocation with lower overall cost than the PT perspective.\\
But the key question is whether the additional expenditure in cybersecurity controls incurred under PT, could yield  a proportionate reduction in risk? To answer this question, we calculate the extent of additional expenditure on $C_{cs}$ and the resulting reduction in $\pi(C_{cs})$ incurred for the baseline PT perspective when compared against the two EUT perspectives.\\
For example, for the risk curve $\pi_1$ and for the under-insurance case $i_r=0.8,$ (see Table \ref{tab:pi1_1}) the investment in $C_{cs}$ under the baseline PT perspective is $16.14,$ which is $10.1\%$ higher than the $C_{cs}$ investment of $14.66$ under both of the EUT perspectives. However, $\pi_1(16.14)$ is $35.8\%$ lower than $\pi_1(14.66).$ This means that in this case, a decision-maker will allocate $10.1\%$ more resources to $C_{cs}$ under the PT perspective compared to the EUT perspective, but in doing so, will also reduce the probability of successful cyber attack by $35.8\%.$ We report the findings for all cases and all risk curves in Tables \ref{tab:pi1_risk} to \ref{tab:pi5_risk}.

\begin{table}[H]
    \centering
    \begin{tabular}{|c|c|c|c|c|c|c|}
    \hline
       \multirow{2}{*}{$\pi_1$}  &\multicolumn{2}{|c|}{$i_r=0$}&\multicolumn{2}{|c|}{$i_r=0.8$}&\multicolumn{2}{|c|}{$i_r=1$}\\ \cline{2-7}
     & $C_{cs}\uparrow$ & $\pi(C_{cs})\downarrow$  &  $C_{cs}\uparrow$ & $\pi(C_{cs})\downarrow$ &  $C_{cs}\uparrow$ & $\pi(C_{cs})\downarrow$ \\ \hline
       vs. EUT ($r=0.88$) & $43.1\%$ & $83.6\%$ & $10.1\%$ & $35.8\%$ & $-0.9\%$ & $-3.9\%$\\ \hline
         vs. EUT ($r=1$) & $43.3\%$ & $83.7\%$ & $10.1\%$ & $35.9\%$ & $-0.9\%$ & $-3.9\%$\\ \hline
      \end{tabular}
    \caption{Risk-reduction and $C_{cs}$ overspend for PT vs. each of the EUT strategies for $\pi_1.$}
    \label{tab:pi1_risk}
\end{table}

\begin{table}[H]
    \centering
    \begin{tabular}{|c|c|c|c|c|c|c|}
    \hline
       \multirow{2}{*}{$\pi_2$}  &\multicolumn{2}{|c|}{$i_r=0$}&\multicolumn{2}{|c|}{$i_r=0.8$}&\multicolumn{2}{|c|}{$i_r=1$}\\ \cline{2-7}
      & $C_{cs}\uparrow$ & $\pi(C_{cs})\downarrow$  &  $C_{cs}\uparrow$ & $\pi(C_{cs})\downarrow$ &  $C_{cs}\uparrow$ & $\pi(C_{cs})\downarrow$ \\ \hline
       vs. EUT ($r=0.88$) & $42.8\%$ & $89.7\%$ & $12\%$ & $48.5\%$ & $-0.4\%$ & $-2.4\%$\\ \hline
         vs. EUT ($r=1$) & $43\%$ & $89.8\%$ & $12\%$ & $48.5\%$ & $-0.4\%$ & $-2.4\%$\\ \hline
      \end{tabular}
    \caption{Risk-reduction and $C_{cs}$ overspend for PT vs. each of the EUT strategies for $\pi_2.$}
    \label{tab:pi2_risk}
\end{table}

\begin{table}[H]
    \centering
    \begin{tabular}{|c|c|c|c|c|c|c|}
    \hline
       \multirow{2}{*}{$\pi_3$}  &\multicolumn{2}{|c|}{$i_r=0$}&\multicolumn{2}{|c|}{$i_r=0.8$}&\multicolumn{2}{|c|}{$i_r=1$}\\ \cline{2-7}
      & $C_{cs}\uparrow$ & $\pi(C_{cs})\downarrow$  &  $C_{cs}\uparrow$ & $\pi(C_{cs})\downarrow$ &  $C_{cs}\uparrow$ & $\pi(C_{cs})\downarrow$ \\ \hline
       vs. EUT ($r=0.88$) & $38.5\%$ & $83.8\%$ & $9\%$ & $35.9\%$ & $-0.9\%$ & $-4.3\%$\\ \hline
         vs. EUT ($r=1$) & $38.7\%$ & $83.9\%$ & $9\%$ & $35.9\%$ & $-0.9\%$ & $-4.3\%$\\ \hline
      \end{tabular}
    \caption{Risk-reduction and $C_{cs}$ overspend for PT vs. each of the EUT strategies for $\pi_3.$}
    \label{tab:pi3_risk}
\end{table}

\begin{table}[H]
    \centering
    \begin{tabular}{|c|c|c|c|c|c|c|}
    \hline
       \multirow{2}{*}{$\pi_4$}  &\multicolumn{2}{|c|}{$i_r=0$}&\multicolumn{2}{|c|}{$i_r=0.8$}&\multicolumn{2}{|c|}{$i_r=1$}\\ \cline{2-7}
        & $C_{cs}\uparrow$ & $\pi(C_{cs})\downarrow$  &  $C_{cs}\uparrow$ & $\pi(C_{cs})\downarrow$ &  $C_{cs}\uparrow$ & $\pi(C_{cs})\downarrow$ \\ \hline
       vs. EUT ($r=0.88$) & $42.5\%$ & $78\%$ & $8.7\%$ & $27.9\%$ & $-1.3\%$ & $-5\%$\\ \hline
         vs. EUT ($r=1$) & $42.7\%$ & $78.1\%$ & $8.7\%$ & $27.9\%$ & $-1.3\%$ & $-5\%$\\ \hline
      \end{tabular}
    \caption{Risk-reduction and $C_{cs}$ overspend for PT vs. each of the EUT strategies for $\pi_4.$}
    \label{tab:pi4_risk}
\end{table}

\begin{table}[H]
    \centering
    \begin{tabular}{|c|c|c|c|c|c|c|}
    \hline
       \multirow{2}{*}{$\pi_5$}  &\multicolumn{2}{|c|}{$i_r=0$}&\multicolumn{2}{|c|}{$i_r=0.8$}&\multicolumn{2}{|c|}{$i_r=1$}\\ \cline{2-7}
       & $C_{cs}\uparrow$ & $\pi(C_{cs})\downarrow$  &  $C_{cs}\uparrow$ & $\pi(C_{cs})\downarrow$ &  $C_{cs}\uparrow$ & $\pi(C_{cs})\downarrow$ \\ \hline
       vs. EUT ($r=0.88$) & $0\%$ & $0\%$ & $0\%$ & $0\%$ & $0\%$ & $0\%$\\ \hline
         vs. EUT ($r=1$) & $0\%$ & $0\%$ & $0\%$ & $0\%$ & $0\%$ & $0\%$\\ \hline
      \end{tabular}
    \caption{Risk-reduction and $C_{cs}$ overspend for PT vs. each of the EUT strategies for $\pi_5.$}
    \label{tab:pi5_risk}
\end{table}

The results show that for all risk curves except $\pi_5,$ the higher allocation of resources to $C_{cs}$ under PT (compared to EUT) seems to reduce $\pi(C_{cs})$ significantly compared to the two EUT allocations for the cases of under insurance and no insurance. For these two cases, the much higher reduction in risk could be considered to be worth the additional expenditure in cybersecurity controls if reducing cyber risk is the key goal. For instance, for the no-insurance case, the PT results in about  $40\%$ additional expenditure on $C_{cs}$ but reduces the probability of a successful attack by upto $80-90\%,$ that is twice as much. For the under-insurance ($i_r=0.8$) the proportionate reduction is even higher at three to four times the proportionate over spend. Thus, PT leads to a higher investment in cybersecurity controls that leads to a proportionately greater reduction in risk.\\
For the full insurance case though, the picture is different. As proved in Theorems 1 and 2, the optimal allocation of resources to $C_{cs}$ under PT will always be less than that under EUT.  Tables \ref{tab:pi1_risk} to \ref{tab:pi5_risk} show that for the case of full insurance, a decision-maker would allocate about $1\%$ less to $C_{cs}$ under PT compared to that under the EUT perspectives, but the $\pi(C_{cs})$ achieved under PT will be about $4-5\%$ higher than that achieved under EUT. Thus, EUT provides a better risk reduction than PT for the full insurance case.

\subsection{Risk curve matters} \label{sec:observations}
Most of the general trends described so far do not hold for the multi-step risk curve $\pi_5.$ For this risk curve, optimal solutions by PT and EUT match exactly for all three cases of no-insurance, under insurance and full insurance. This highlights that understanding the nature of the risk curve is critical in understanding how different approaches to resource allocation will fare.\\
The shape of the risk curve is really important. In particular, we observe the following.
\begin{itemize}
\item{\emph{How quickly the risk reduces:} Both security and insurance costs are much higher when the risk reduces slowly with investment in security controls ($\pi_1$ and $\pi_3$) for both PT and EUT.}
\item{\emph{Baseline risk is high:} Where the baseline risk is higher ($\pi_3$), the decision-makers that behave in line with PT or EUT tend to invest more in cybersecurity controls but we see differing costs invested in cyber insurance depending on how quickly the risk reduces with further investment.}
\item{\emph{Step function:} Where the rate of reduction in risk takes the form of step function, we see that both PT and EUT perspectives will lead to investing much more in both the cybersecurity controls as well as the cyber insurance! Further, for the multi-step risk curves, there may not be any difference in the resource allocation obtained under either of the two perspectives!}
\end{itemize}

\subsection{Finding the optimal allocation for an organization}

If an organization was given the three insurance options discussed, which option will be the most favored if they acted in line with the PT perspective versus the EUT perspective? We find that regardless of the risk curve, when acting in line with PT, the decision-maker will always see the full insurance case as the best option (highest overall value). Whereas when acting in line with either of the EUT perspectives they'd find the no insurance case to be the best option (highest expected utility).\\
However, if the choice to go with the full insurance option was already made, then considering EUT would yield a better resource allocation than considering PT since it seems to provide better risk reduction for the same total cost.\\
On the other hand, if under insurance was the only option (as can often be the case, in practice) then the resource allocation under PT could provide superior risk reduction but this could come at a higher cost than the resource allocation under EUT. As discussed, this cost could be substantially higher if the insurance coverage was lower.\\
So far we have compared the implications of using the EUT perspective versus the PT perspective on the optimal resource allocation to cybersecurity controls and cyber insurance for different risk curves. However, now we we will briefly consider a different decision-making problem.\\
\noindent \textbf{Conscious attempt to create a bespoke strategy}\\
Suppose that an organization has established the shape of their risk curve and concluded that under-insurance is the only feasible case. The organization wants to find the \emph{best of the both worlds} resource allocation. That is, an allocation that results in superior risk reduction than achieved under an EUT perspective, but without the additional cost incurred under the PT perspective.  So, given a risk curve, is it possible to find the optimal combination of PT parameters $\alpha$ and $\beta$ so as to achieve superior risk reduction than EUT but at no extra cost? Our conjectures and numerical results indicate that this should be possible.\\
For instance, consider the case of $i_r=0.8$ for risk curve $\pi_1.$ From Table \ref{tab:pi1_1}, we can see that the total cost for EUT, $r=1$ is $17.46$ of which $14.66$ is allocated to $C_{cs}.$ Then using a simple grid search algorithm, we find that using PT with $\alpha=0.97$ and $\beta = 0.97$ will yield the same total allocation of $17.46,$ of which $14.71$ ($0.3\%$ higher) is for $C_{cs}$ resulting in $\pi(C_{cs})$ that is $1.6\%$ lower, thus achieving a slightly better risk reduction than EUT for the exact same total cost. 

\subsection{Sensitivity Analysis}

Since the scale of operations will be different for different organizations, so will be their wealth, losses and cybersecurity control expenditures. From the optimal resource allocation point of view, the relative proportions of these quantities is important, not the absolute values. In our examples, we have assumed that the losses are $10\%$ of the wealth and the cybersecurity control expenses are only a fraction of the overall wealth, which we believe is realistic. In fact, for the losses, we have assumed a much higher proportion of losses as a worst case scenario than the under $4\%$ reported by current research \citep{Kamiya2021,Walton2021}.\\
We performed sensitivity analysis to understand the robustness of our numerical
results and found that the key findings are robust not only to varying values of $W,$ but also to different relative proportions of $L$ with respect to $W.$ We also found that they are robust to changes in profit margin $q,$ including the extreme case of fair premiums $q=0.$

\section{Conclusion and further work}
\subsection{Key findings}
To the best of our knowledge, ours is the first piece of research that considers the optimal resource allocation to cybersecurity controls and cyber insurance using PT. More importantly, we compare and contrast the optimal allocation obtained using PT and EUT. Using conjectures, mathematical results and numerical results, we illuminate why PT could lead to different resource allocation than EUT. We also introduce the concept of risk curves and our results show that having an accurate understanding of the shape of the risk curve is critical in correctly deriving the optimal resource allocation.\\ 
PT has been proposed and widely discussed as a \emph{descriptive approach} to better understand how most people would make decisions in risky situations. However, our work shows that there maybe merit in conscious use of PT as a decision-making strategy by experts in professional resource allocation settings. Which of the two approaches is more appropriate depends on the key objective that the organization is aiming to achieve, namely, is it risk reduction or cost optimization? It might even be possible, subject to the nature of the risk curve, to create a bespoke strategy that achieves both.\\
We believe that our work not only provides a practical framework for organizations to allocate resources to cybersecurity but could also help regulators and policy makers in developing more informed advice and policies around cybersecurity resource allocation. 

\subsection{Limitations} 
Our results are only limited to EUT using a standard CRRA utility function. 
 Risk aversion can be modeled in a utility function using many other ways \citep{Back2017} and other behavioral characteristics such as regret can also be considered in a utility function \citep{ejaz22}. 
It is beyond the scope of this manuscript to perform an exhaustive comparison that considers all possible variations of EUT and PT.

\subsection {Further Work}
We do not discuss how to derive the risk curves. The risk curves used in this study have been modeled using exponential decay functions that we believe would emulate the differing risk profiles that could exist in practice. However, it is not obvious how to infer the true risk curve for a give situation. As discussed in Section 1, cyber risk as well as the effectiveness of cybersecurity controls are challenging to quantify. Further research on accurately quantifying risk curves for cyber risk is needed.

\bibliographystyle{apalike}
\bibliography{Cybersec_resource_allocation_Joshi_etal}
\end{document}